\documentclass[useAMS,usenatbib]{mn2e}

\usepackage{graphicx}
\usepackage{amsmath}
\usepackage{float}

\title[Hurst parameter of pulsar timing noise]{Hurst parameter analysis of radio pulsar timing noise}

\author[Na et al.]
{X. S. Na$^{1}$, N. Wang$^{2}$, J. P. Yuan$^{2}$, Z. Y. Liu$^{2}$, J. Pan$^{3}$ and R. X. Xu$^{1}$\\
$^1$School of Physics and State Key Laboratory of Nuclear
Physics and Technology, Peking University, Beijing 100871, China\\
$^2$Urumqi Observatory, National Astronomical Observatories,
40-5 South Beijing Road, Urumqi, 830011, China\\
$^3$Purple Mountain Observatory, 2 West Beijing Road, Nanjing 210008, China \\
}

\begin{document}
\maketitle
\begin{abstract}
We present an analysis of timing residual (noise) of $54$ pulsars
obtained from 25-m radio telescope at Urumqi Observatory with a time
span of $5\sim 8$ years, dealing with statistics of the Hurst
parameter. The majority of these pulsars were selected to have
timing noise that look like white noise rather than smooth curves.
The results are compared with artificial series of different
constant pairwise covariances. Despite the noise like appearance,
many timing residual series showed Hurst parameters significantly
deviated from that of independent series. We concluded that Hurst
parameter may be capable of detecting dependence in timing residual
and of distinguishing  chaotic behavior from random processes.
\end{abstract}

\begin{keywords}
pulsars: general -- methods: statistical
\end{keywords}
\section{Introduction}\label{sec_intro}
All pulsars show a remarkable uniformity of rotation rate on a time
scale of a few days as expected of an isolated spinning body with
large stable moment of inertia. \citep{Lyne06} The angular momentum
of radio pulsar is slowly decreasing through slowdown torque of the
magnetic dipole radiation. However, some very interesting
irregularities in pulsar rotation have been observed which are
termed as {\em timing noise}.

It is anticipated that valuable information of many interesting
physical processes related to pulsars is coded in the timing noises,
thus employing statistical measures to characterize timing noises is
important to the study of pulsars and thus the properties of the
matter at supra-nuclear densities. Efforts to quantify timing noise
have been tried as early as timing noise was firstly recognized, for
instance according to random walk of different quantities
\citep{Boynton72}, most current models such as vortex creeping are
still restricted to treatment of timing noise only as random process
in certain quantities. One exception was presented by
\citep{Harding90}, who analyzed timing data of Vela pulsar to look
for evidence of chaotic behavior by ``correlation sum'' technique to
estimate fractal dimension of the system. However, despite possible
suggestions they concluded that ``correlation sum'' estimator may
be unable to distinguish between random and chaotic processes.

However, any statistical representation of data has their own
biases, employing a large types of statistical measures is
essentially vital to a fair understanding of the timing noises.
Furthermore, the number of observed pulsars has accumulated to $\sim
10^3$ \citep{Manchester06} but collected data is often incomplete
for a conclusive analysis, it is therefore crucial to diagnose
current available but limited data, the results of which could guide
us in future observation to concentrate on those pulsars with
anomalous timing noise. Here we are introducing a statistical method
rarely used in time domain astronomy -- the Hurst parameter
analysis, which is actually sensitive to the type of the inherent
correlation among the time series. Our practical analysis of the
timing data observed by the 25-meter radio telescope at Urumqi
Observatory of $54$ pulsars
indicates that Hurst parameter analysis might be capable of
detecting anomalous signals which disguise themselves as noises.

\section{The Hurst parameter analysis}\label{sec_2}
\subsection{Basic concepts}\label{sec_concept}
The timing noise in essence is a discrete realization of a continuous
random process $\{ X_t \}$, let $\{X_1, X_2, \ldots, X_n,
\ldots\}$ be a discrete time series sampled sequentially at time
points $t_1, t_2, \ldots, t_n, \ldots$ and the task is to analyze
the record $\{ X_n \}$ and identify corresponding characteristics of
$\{ X_t \}$.

The first property one may check is whether there is correlation
among the series. The most commonplace time series is the
\emph{independently distributed random series} if the covariance
$\mbox{Cov}(X_i,X_j)=0$ for any $i\neq j$. The independently
distributed random series is also the well known as {\em white
noise} for its constant power spectrum. A closely related concept is
the colored noises which are called so for their power-law shaped
power spectrum $P(k)\propto k^n$ ($n\ne 0$).

Another important aspect is stationarity.
Stationary means that for any $d$ indices $k_1,\ldots,k_d$ the
vectors $(X_{k_1},\ldots,X_{k_d})$ and $(X_{k_1+n},\ldots,X_{k_d+n})$
have the same n-point distribution \citep{Dieker04}. To ensure the
stationarity, one can either enlarge the observation window to have
sufficiently long record of data to include the short-term variation,
or construct a new set of data by de-trending the observed time
series with appropriate models.

If for the new series that sums $m$ consecutive variable
$Y_k=X_{km}+X_{km+1}+\ldots+X_{km+m-1}$, there exists a scaling
function $a(m)$ so that for any $d$ indices $k_1,\ldots,k_d$ the
vectors $(X_{k_1},\ldots,X_{k_d})$ and $a(m)(Y_{k_1},\ldots,Y_{k_d})$
have the same n-point distribution, we call the time series
\emph{self-similar}. \citep{Dieker04} Self-similar time series thus
looks the same after scaling up to a factor $a(m)$ while $m$ plays the
role of resolution. If for self-similar series the function $a(m)$
obeys power law $a(m)\propto m^{-H}$, we call the series a
self-similar series with a \emph{Hurst parameter} of $H$.

It is obvious that an independently and identically
distributed\footnote{ All random variables within the series have
the same distribution and are independent with one another,
abbreviated as i.i.d.} Gaussian series is self-similar with Hurst
parameter $H=1/2$. But once the series has some non-trivial correlation
structures, the Hurst parameter may differ from $1/2$, which is better
illustrated with the following introduction of Hurst parameter
\citep{Feder1988} for continuous-time stochastic process.
If $X(t)$ is a random process on some probability space such that:
\begin{enumerate}
\item with probability 1, $X(t)$ is continuous and $X(0)=0$;
\item for any $t\ge 0$ and $\tau > 0$, the increment $X(t+\tau)-X(t)$ follows
a normal distribution with mean zero and variance $\tau^{2H}$, so
that
\begin{equation}
P\left( X(t+\tau)-X(t)\le x
\right)=\frac{\tau^{-H}}{\sqrt{2\pi}}\int_{-\infty}^x
e^{-u^2/2\tau^{2H} }du\ . \label{eq:hdef}
\end{equation}
\end{enumerate}
The parameter $H$ is then the Hurst parameter (or Hurst exponent),
if $H=1/2$, the random process is just the i.i.d Gaussian random
process \citep[c.f.][]{Feder1988}. It is easy to see that the
correlation
\begin{equation}
\langle X(t) \left[X(t+\tau)-X(t)\right] \rangle = \frac{ (
t+\tau)^{2\alpha}-t^{2\alpha}-\tau^{2\alpha}}{2} \label{eq:corr}
\end{equation}
does not vanish unless $H=1/2$.

Random walk can be formed from a random process $X$ via $\int X(t)
dt$ or $\sum_i X_i$, the random walk with pace length drawn from
i.i.d Gaussian series is the normal Brownian motion while those
walks coming from Eq.~\ref{eq:hdef} with $H\in (0, 1)$ and $H\ne
1/2$ is named {\em fractional Brownian motion}. As can be seen from
Eq.\ref{eq:corr}, a fractional Brownian motion with Hurst parameter
$H>1/2$ contains a positive correlation between consecutive steps or
in other words a persistent trend, and covers wider range
than normal Brownian motion~\citep{Steeb05}. For series with $H<1/2$
the generated random walk will cover less than Brownian motion,
implying a negative correlation between consecutive variables or in
other words a self-reverting trend, i.e. being
\emph{anti-persistent}. Therefore, estimation of Hurst parameter can
probe correlation inside the series, i.e. how the `memory' of its
past affects the future. The method was named after its proposer H.
Hurst~\citep{Hurst51}, and has been widely used in various areas
such as biology, medical science, seismology, and economics.

\subsection{Estimation of the Hurst exponent}\label{sec_est}
Our algorithm is a slightly modified version of the popular Range
Over Standard deviation (ROS) algorithm ~\citep{Steeb05} to account
for the non-uniform sampling\footnote{We divide the series into segments
of roughly equal time span instead of equal number of data points because usually radio pulsar
TOA measurements are not uniform.}. The timing noise record
$X_1,X_2,\ldots,X_N$ taken at time $t_1,\ldots,t_N$, is firstly
split into $n$ segments of length $\Delta t_n\approx (t_N-t_1)/n$,
i.e. $\{ X_N\}$ is re-grouped in the way
\begin{equation*}
X_1,\ldots,X_{\ell_1}; X_{\ell_1+1},\ldots,X_{\ell_2}; \quad \ldots
\quad ; X_{\ell_{n-1}+1},\ldots,X_N
\end{equation*}
so that
\begin{equation}
\label{eq:001} t_{\ell_i},t_{\ell_i+1},\ldots,t_{\ell_{i+1}-1}\in
[t_1+(i-1)\Delta t,t_1+i\Delta t)
\end{equation}
For $i$-th segment $\{X_{\ell_i},\ldots,X_{\ell_{i+1}-1}\}$ for
example, we calculate average $A_i$ and standard deviation $S_i$ of
the record
$\{0,X_{\ell_i+1}-X_{\ell_i},\ldots,X_{\ell_{i+1}-1}-X_{\ell_i}\}$
to form a new series $\{Y_{\ell_i},\ldots,Y_{\ell_{i+1}-1}\}$ whose
element is given by
\begin{equation}
Y_{\ell_i+k}=\frac{1}{S_i}\left(X_{\ell_i+k}-A_i\right)\ .
\end{equation}
The \emph{range} $R_i^{(n)}$ is defined to be the difference between
the maximum and the minimum of the accumulated series,
\begin{equation}
\begin{aligned}
Z^{(i)}_{k}&=\sum_{j=1}^{k}Y_{\ell_i+j}\quad (k\leq \ell_{i+1}-\ell_i-1)\\
R_i^{(n)}&\equiv {\rm Max}(Z^{(i)}_{k=1, \ldots}) -{\rm
Min}(Z^{(i)}_{k=1, \ldots})
\end{aligned}
\end{equation}
where superscript $n$ is the number of segments, and for each $n$
there is an averaged range
\begin{equation}
R^{(n)}=\frac{1}{n}\sum_{i=1}^n R_i^{(n)}\ .
\end{equation}
Therefore for $n=1,2,\ldots,n_{\mbox{max}}$ we obtain a sequence
$\{(\Delta t_n,R^{(n)})\}$. $n_{\mbox{max}}$ is chosen so to be the
largest segment number to ensure that every segment contain at least
$5$ data points.
To estimate the power-law index we then use linear regression to fit
$\{(\log\Delta t_n,\log R^{(n)})\}$ for the slope as estimation
of the Hurst parameter $H$.

\subsection{Simple Monte-Carlo simulation}\label{sec_mc}
However there two complications in practical estimation of  Hurst
parameter of timing residual data:

\begin{enumerate}
\item The algorithm is for series of exact values while for timing noise looks like white noise, uncertainties are
usually not negligible.
\item The algorithm uses running maximum while the definition uses variance. These two should show
the same trend in $n\to \infty$ limit \citep{Karatza91} but there is
no information to what extend this is true for finite sequence.
\end{enumerate}

In order to control possible systematical effects, we carry out
numerical test with simple simulations. \footnote{In all Monte-Carlo
simulations in this article we used Mersenne Twister with a period
of $2^{19937}-1$ for pseudo-random number generator~\citep{Matsumoto98}.}
We generate $10^3$ series $\{(t_n,X_n)\}$ for each pulsar to take
error bar effect into account where $n$ labels data points number and
in each sequence $X_n$ follows the Gaussian distribution with standard
deviation equal to data error bar lengths and expectation equal to the
centers of data error bar. Then Hurst parameters $H$ of these series
and their average and standard deviation is calculated for each pulsar.

Next we generate artificial (anti-)persistent series $\{(t_n,Y^{(c)}_n)\}$
with constant covariance $c$ between consecutive pair at observation
times $t_n$ in order to control systematic error from finite length.
In other words, we generate a Gaussian random number $Y^{(c)}_1\sim N(0,1)$
at $t_1$, then with the value of $Y^{(c)}_1$ we generate independent random
number $Y^{(c)}_2\sim N(cY^{(c)}_1,1)$ for $t_2$, and so forth for the rest
of $t_{n}$ there is the $Y^{(c)}_n\sim N(c Y^{(c)}_{n-1},1)$ where $c$ is
clearly the constant covariance between consecutive numbers. Then we estimate
the Hurst parameter $H^{(c)}$ for each of these series $\{Y^{(c)}\}$.
These values would serve as a standard for us to compare with $H$ obtained above.

Note that these series is self-similar but do not have power-law
scaling function and therefore do not have a strictly defined Hurst
parameter. This can be shown by a simple calculation. Given an
i.i.d. series with standard Gaussian distribution
$(X_1,X_2,\ldots,X_n,\ldots)$, the process with covariance $c$ can
be expressed as
\begin{equation*}
(Y_1=X_1,Y_2=cX_1+X_2,\ldots,Y_k=\sum_{k=1}^n c^{n-k}X_k,\ldots)
\end{equation*}
and variance of sum of first $n$ variables is
\begin{equation}
\label{eq:in01}
\mbox{Var}\left(\sum_{i=1}^nY_i\right)=\sum_{m=0}^{n-1}\left(\sum_{k=0}^m
c^k\right)^2=\alpha c^{2n+1}+\beta c^{n+1}+\gamma n+\delta
\end{equation}
with
\begin{equation}
\begin{cases}
\alpha&=-\frac{1}{(1+c)(1-c)^3}\\
\beta&=\frac{2}{(1-c)^3}\\
\gamma&=\frac{1}{(1-c)^2}\\
\delta&=-\frac{c(c+2)}{(1+c)(1-c)^3}\ .
\end{cases}
\label{eq:in02}
\end{equation}
Hence its scaling function is
\begin{equation}
a(n)^2=\frac{1}{\alpha c^{2n+1}+\beta c^{n+1}+\gamma n+\delta}
\label{eq:008}
\end{equation}
Clearly it is not power-law except for $c=0$ which leads to
$a(n)\propto n^{-1/2}$. However, we can still use
linear regression to find slope $-H$ of the log-log curve of $a(n)$
which gives $H=0.43,0.47,0.56,0.74$ for $c=-0.8,-0.4,0.4,0.8$,
$n=1\sim 100$, and $H=0.45,0.48,0.54,0.67$ for $c=-0.8,-0.4,0.4,0.8$ ,
$n=1\sim 200$. Thus these series can simulate (anti-)persistent
series with short length. For long length series with $c<1$, $a(n)$
will quickly approach $n^{-1/2}$ as $n\to\infty$, in other word
$H=1/2$ in the large $n$ limit. This simply calculation also demonstrates
that Hurst parameter deviation from $1/2$ seen in long series cannot
be attributed only to covariance between only consecutive pairs but reveals
long-range dependence which is an essential feature of fractional
Brownian motion.

\section{Practical Data Analysis}\label{sec_prac}
\subsection{The sample of observed timing noises}\label{sec_datas}
Data used in this work is $5\sim 8$ years of timing residual data
from 25m radio telescope at Urumqi Observatory. Timing data is processed using
standard pulsar software packages PSRCHIVE \citep{Hotan04} and TEMPO2 \citep{Hobbs06}
to obtain timing residual. Time of arrival (TOA) of radio pulse is fitted
using PSRCHIVE software from observation data by an input accumulated pulse
profile. TEMPO2 software use coordinate system conversion parameters that
account for various effects (Roemer delay, Einstein delay, Shapiro delay etc.)
and pulsar parameters ($P$, $\dot{P}$ etc.) to establish a model to predict
arrival time of pulses. Timing residual is obtained as difference between
observation and this timing model. We then use splk package of TEMPO2 to output
timing residual with error bar in unit of second at each TOA. We select $54$ pulsars
among pulsars regularly monitored at Nanshan with no intention to be make our
selection statistically representative. Because of the nature of our analysis
method, we mainly select pulsars with small $\dot{P}$ and timing noise like
white-noise without recognizable glitch. A list of these pulsars with various parameters
($P$, $\dot{P}$, dispersion measure and characteristic age $\tau_c$) from the ATNF
Pulsar Database~\citep{Manchester06} is presented Table \ref{tab_ATNF}.

\subsection{Results and analysis}\label{sec_result}
\begin{figure}
\includegraphics[scale=1]{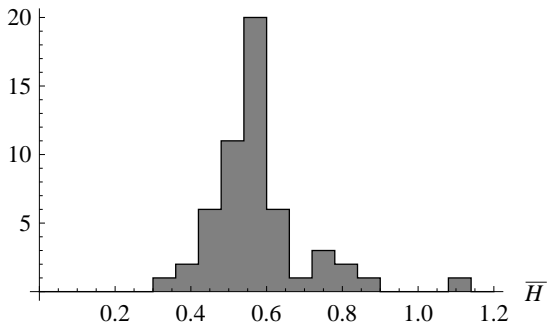}
\centering
\caption{Distribution of Hurst parameters for $54$
pulsars. All $H$ values are in the interval $0\sim 1.2$.}
\label{fig_1}
\end{figure}

\begin{figure}
\includegraphics[scale=0.4]{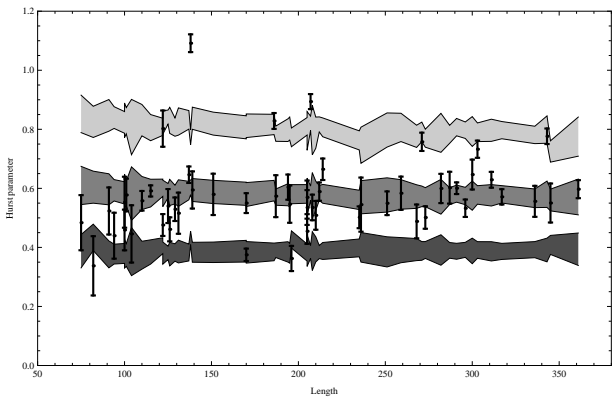}
\centering
\caption{Hurst parameter of $54$ pulsars. The black
error bar shows average and standard deviation of $1000$ Hurst
parameter value for each pulsar. In comparison artificial
(anti-)persistent series with constant covariance $c=-0.8,0,0.8$ are
plotted as bands of connected error bars represented by increasing
gray level.}
\label{fig_2}
\end{figure}

Using the above methods, Hurst parameters $\{H\}$ and
$\{H^{(c)}\}$ with $c=-0.8,-0.4,0,0.4,0.8$ for $54$ radio
pulsars are calculated.
Average values and standard deviations of these two groups of Hurst
parameters are listed in Table \ref{tab_H} together with basic
parameters of $54$ pulsars. Distribution of average $\bar{H}$ is
shown in Figure \ref{fig_1}. In Figure \ref{fig_2}, we plot average value and standard
deviation of $\{H\}$ for each pulsar as a black error bar and in
comparison average and standard deviation of $\{H^{(c)}\}$ for
artificial (anti-)persistent series with $c=\pm 0.8$ and i.i.d.
series with $c=0$ are plotted as connected error bar bands. For
clarity we do not show bands of artificial (anti-)persistent series
with $c=\pm 0.4$ which roughly fall in the gaps between $c=0$ and
$c=\pm 0.8$ but have some overlap.

From the average and standard deviation of Hurst parameters of the
three bands shown in Figure \ref{fig_2}, we see that data length and
non-uniformity has only limited effect on the Hurst parameter
estimation. The values of Hurst parameters of the three groups of
artificial series are not exactly the same with Hurst parameters
calculated using variance in \S \ref{sec_mc}. This can be attributed to the
difference between calculation by definition and our algorithm using
running maximum as mentioned in \S \ref{sec_mc}.

From these results, we can see that most Hurst parameters for pulsar
timing noises concentrate around $0.5$. However there are a small
portion with $H$ values far away from $0.5$ despite considerable
data point number. For PSRs J0147$+$5922, J0357$+$5236,
J0612$+$3721, J0630$-$2834, J0823$+$0159 and J0837$-$4135 calculated
Hurst parameter average $H>0.75$ and as is shown in Figure \ref{fig_2}
their error bars appear together with or even above the $c=0.8$
band, indicate that they have strong persistent trend. On the other
hand, for PSRs J0055$+$5117, J1022$+$1001, and J1842$-$0359
calculated Hurst parameter $H<0.4$ and as is shown in Figure \ref{fig_2} their
error bars appear together with or even below the $c=-0.8$ band,
indicating strong anti-persistent trend.

\begin{figure}
\includegraphics[scale=0.9]{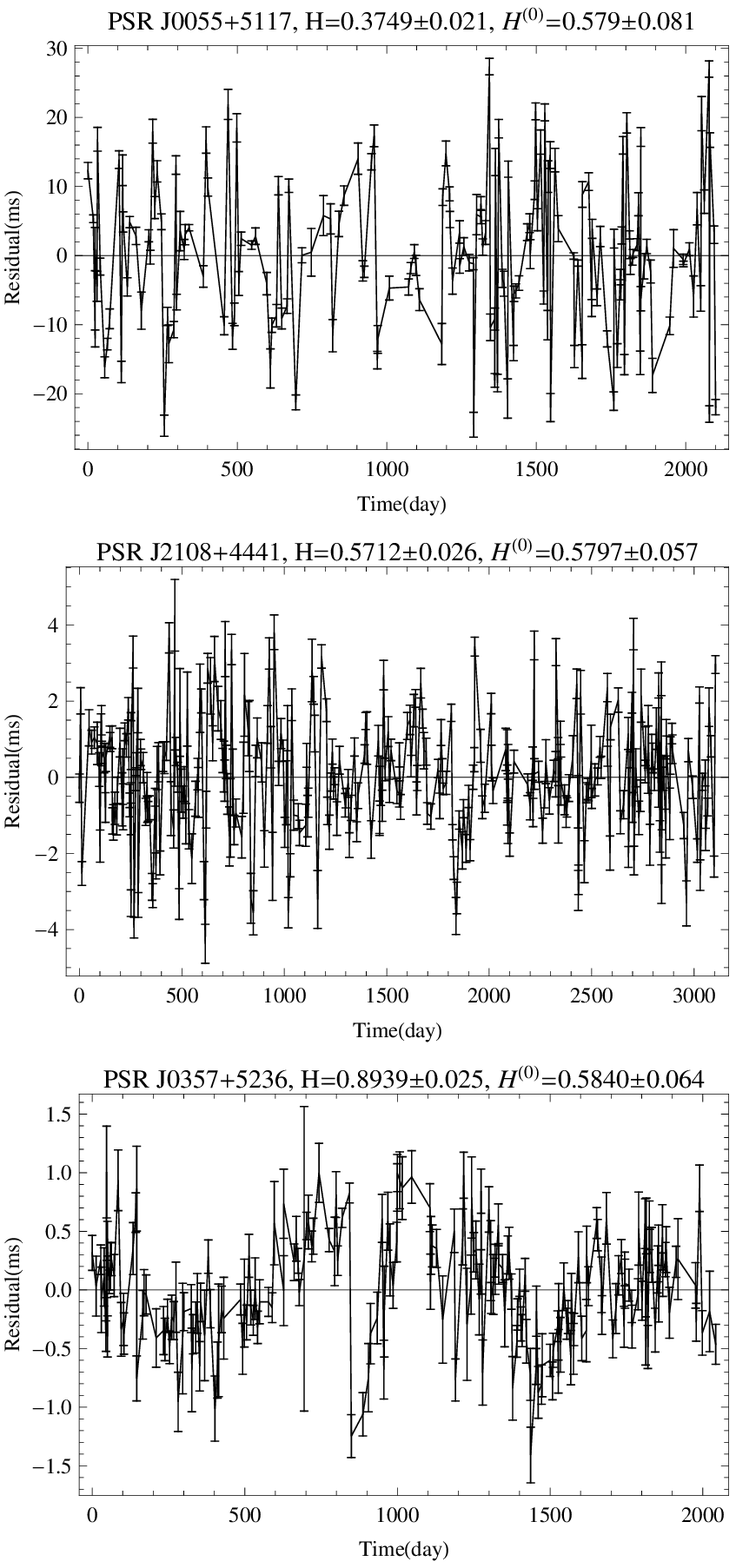}
\centering
\caption{Representative timing noises. PSR J0055$+$5117
representing anti-persistent series ({\it top panel}), PSR
J2108$+$4441 representing independent series ({\it middle panel}),
and PSR J0357$+$5236 representing persistent series ({\it bottom
panel})}
\label{fig_3}
\end{figure}

To illustrate the typical persistent series and anti-persistent
series seen in real pulsar timing noise, we pick three pulsars: PSR
J0055$+$5117 with $H=0.3749\pm 0.021$ and $H^{(0)}=0.5790\pm 0.081$
representing anti-persistent series, PSR J2108$+$4441 with
$H=0.5712\pm 0.026$ and $H^{(0)}=0.5797\pm 0.057$ representing
independent series and PSR J0357$+$5236 with $H=0.8939\pm 0.025$ and
$H^{(0)}=0.5840\pm 0.064$ representing persistent series. Their
timing noises in unit of millisecond are plotted in Figure \ref{fig_3}.

\section{Discussion and Conclusion}\label{sec_cncl}
We calculated Hurst parameter of $54$ radio pulsars with
white-noise-like timing residual obtained from Nanshan telescope and
compared the results with artificial (anti-)persistent series. The
majority of pulsars from our selection have Hurst parameters around
$0.5$ and not far from Hurst parameter calculated for independent
series. However, we found 9 pulsars (PSRs J0147$+$5922,
J0357$+$5236, J0612$+$3721, J0630$-$2834, J0823$+$0159 and
J0837$-$4135 showing persistent trend and PSRs J0055$+$5117,
J1022$+$1001, and J1842$-$0359 showing anti-persistent trend) with
interesting $H$ values despite having white-noise-like timing
residual. Comparison with artificial series confirm that these
trends cannot be attributed entirely to effects caused by ROS
algorithm or large uncertainty in timing residual. This shows that
our algorithm may be capable of detecting hidden correlation
in apparently noise-like timing residual. We therefore suggest that these 9
pulsars be monitored continuously to confirm or disprove long-range
dependence and search for possible physical process behind
such correlation.

As for the selection, we picked $54$ pulsars with relatively small
period derivative $\dot{P}$, and white-noise-like timing residual
rather than the typical red-noise pattern of the three kind of noise
model (PN, FN and SN) first proposed by Boynton et al. (1972) and of
course we made no attempt at doing correlation between Hurst
parameter and other pulsar properties nor obtaining statistics of
Hurst parameter values for large population. Our method is aimed at
finding possibility that timing noise that resembles white-noise is
not really generated by a random process. As is well known there are
pulsars with smooth timing noise that differ largely from white
noise (such as J0332+5434, J0406+6138, J0826+2637, etc.) and they
would yields quite large Hurst parameter if calculated using our
algorithm and some even exceed $1$. In that case we do not need the
deviation from $0.5$ to conclude the obvious dependent nature of
timing noise and other method should be used to analyze possible
chaotic nature of these timing noise series.

There are various physical processes that might be responsible
for the long-range dependence of pulsar timing noise. These can be
classified into three groups: from interior of neutron star, i.e.
due to fluctuation of internal (e.g. micro-quake due to partial
release of elastic energy \cite{Pines72} and random pinning and
unpinning of vortex lines \cite{Packard72}, \cite{Anderson75}) and
external (e.g. accretion flow \cite{Lamb78}) torques; from emission
process (e.g. magnetosphere activities \cite{Cheng87}) and from
propagation of radio emission. For the last class of origin, it has
long been proposed that timing noise can be used to set upper limits
on gravitational wave (\cite{Bertotti83}) and that given enough time
pulsar timing array might be the first equipment to directly detect
gravitational wave (\cite{Manchester062}). All of these (including
gravitational wave detection which aims at detecting random
background) predict randomness in evolution of certain quantity
while Hurst parameter is capable of uncovering chaotic behavior that
is hidden in timing noise. The physical origin that leads to
dependence series is much more limited than random fluctuations.
Therefore, long-term dependence detected by Hurst parameter might
reveal more detailed information about the physical origin of timing
noise. For instance it has been proposed that under certain
conditions Euler equations for rotating object with magnetic dipole
moment misaligned with rotation axis would show chaotic spin-down
behavior (see \cite{Harding90}). By definition any dynamical process
can exhibit chaotic behavior and thus dependence in long-term timing
noise data if the system is governed by differential equations whose
solution is highly sensitive to initial conditions. We expect that
the application of Hurst parameter to timing noise of longer time
span reveal more evidence for new physics.

\section*{Acknowledgments}
We would like to acknowledge useful discussions at our pulsar group
of PKU. This work is supported by NSFC (10778611), the National
Basic Research Program of China (grant 2009CB824800) and by LCWR
(LHXZ200602).

\appendix

\begin{table*}
\begin{tabular}{|l|lllllll}
\hline
JName      & $P$(s)   & $\dot{P}$($10^{-16}$)  & DM(cm$^{-3}\cdot$pc) & $\tau_c$($10^7$yr) \\
\hline
J0034$-$0721 & $0.94295099$ & $4.0821$ & $11.38$ & $3.66$ \\
J0055$+$5117 & $2.11517114$ & $95.3764$ & $44.13$ & $0.351$ \\
J0134$-$2937 & $0.13696158$ & $0.7837$ & $21.81$ & $2.77$ \\
J0141$+$6009 & $1.22294852$ & $3.9107$ & $34.80$ & $4.95$ \\
J0147$+$5922 & $0.19632127$ & $2.5677$ & $40.11$ & $1.21$ \\
J0151$-$0635 & $1.46466454$ & $4.4259$ & $25.66$ & $5.24$ \\
J0152$-$1637 & $0.83274161$ & $12.992$ & $11.92$ & $1.02$ \\
J0215$+$6218 & $0.54887981$ & $6.6212$ & $84.00$ & $1.31$ \\
J0304$+$1932 & $1.38758444$ & $12.9524$ & $15.74$ & $1.7$ \\
J0357$+$5236 & $0.19703009$ & $4.7659$ & $103.71$ & $0.655$ \\
J0502$+$4654 & $0.63856548$ & $55.8234$ & $42.19$ & $0.181$ \\
J0525$+$1115 & $0.35443759$ & $0.7361$ & $79.35$ & $7.63$ \\
J0601$-$0527 & $0.39596916$ & $13.0210$ & $80.54$ & $0.482$ \\
J0612$+$3721 & $0.29798232$ & $0.5947$ & $27.14$ & $7.94$ \\
J0630$-$2834 & $1.24441859$ & $71.229$ & $34.47$ & $0.277$ \\
J0758$-$1528 & $0.68226517$ & $16.1889$ & $63.33$ & $0.668$ \\
J0814$+$7429 & $1.29224144$ & $1.6811$ & $6.12$ & $12.2$ \\
J0820$-$1350 & $1.23812954$ & $21.0518$ & $40.94$ & $0.932$ \\
J0823$+$0159 & $0.86487280$ & $1.0455$ & $23.73$ & $13.1$ \\
J0837$+$0610 & $1.27376829$ & $67.9922$ & $12.89$ & $0.297$ \\
J0837$-$4135 & $0.75162361$ & $35.3930$ & $147.29$ & $0.336$ \\
J0846$-$3533 & $1.11609716$ & $16.0135$ & $94.16$ & $1.1$ \\
J0908$-$1739 & $0.40162562$ & $6.6950$ & $15.89$ & $0.95$ \\
J0943$+$1631 & $1.08741772$ & $0.9109$ & $20.32$ & $18.9$ \\
J1022$+$1001 & $0.01645292$ & $0.0004$ & $10.25$ & $601$ \\
J1041$-$1942 & $1.38636807$ & $9.4485$ & $33.78$ & $2.32$ \\
J1115$+$5030 & $1.65643975$ & $24.9279$ & $9.20$ & $1.05$ \\
J1239$+$2453 & $1.38244910$ & $9.6005$ & $9.24$ & $2.28$ \\
J1257$-$1027 & $0.61730766$ & $3.6270$ & $29.63$ & $2.7$ \\
J1543$+$0929 & $0.74844841$ & $4.3248$ & $35.24$ & $2.74$ \\
J1703$-$3241 & $1.21178509$ & $6.5983$ & $110.31$ & $2.91$ \\
J1733$-$2228 & $0.87168283$ & $0.4270$ & $41.14$ & $32.3$ \\
J1741$-$0840 & $2.04308245$ & $22.7471$ & $74.90$ & $1.42$ \\
J1750$-$3157 & $0.91036298$ & $1.9652$ & $206.34$ & $7.34$ \\
J1756$-$2435 & $0.67047996$ & $2.8474$ & $367.10$ & $3.73$ \\
J1820$-$1818 & $0.30990459$ & $0.9361$ & $436$ & $5.25$ \\
J1822$-$2256 & $1.87426851$ & $13.5439$ & $121.20$ & $2.19$ \\
J1823$+$0550 & $0.75290654$ & $2.2673$ & $66.78$ & $5.26$ \\
J1834$-$0426 & $0.29010819$ & $0.7195$ & $79.31$ & $6.39$ \\
J1837$-$0653 & $1.90580870$ & $7.7204$ & $316$ & $3.91$ \\
J1840$+$5640 & $1.65286185$ & $14.9482$ & $26.70$ & $1.75$ \\
J1842$-$0359 & $1.83994431$ & $5.0876$ & $195.98$ & $5.73$ \\
J1852$-$2610 & $0.33633713$ & $0.8771$ & $56.81$ & $6.08$ \\
J1900$-$2600 & $0.61220920$ & $2.0453$ & $37.99$ & $4.74$ \\
J1901$-$0906 & $1.78192776$ & $16.3829$ & $72.68$ & $1.72$ \\
J1921$+$2153 & $1.33730216$ & $13.4821$ & $12.46$ & $1.57$ \\
J1946$+$1805 & $0.44061847$ & $0.2409$ & $16.22$ & $29$ \\
J1954$+$2923 & $0.42667678$ & $0.0171$ & $7.93$ & $395$ \\
J2018$+$2839 & $0.55795348$ & $1.4811$ & $14.17$ & $5.97$ \\
J2046$-$0421 & $1.54693811$ & $14.7148$ & $35.80$ & $1.67$ \\
J2046$+$1540 & $1.13828568$ & $1.8232$ & $39.84$ & $9.89$ \\
J2108$+$4441 & $0.41487053$ & $0.8621$ & $139.83$ & $7.62$ \\
J2113$+$4644 & $1.01468479$ & $7.1461$ & $141.26$ & $2.25$ \\
J2308$+$5547 & $0.47506767$ & $1.9949$ & $46.54$ & $3.77$ \\
\hline
\end{tabular}
\caption{Parameters of $54$ pulsars used in Hurst parameter
calculation from ATNF Pulsar Catalog \citep{Manchester06}}
\label{tab_ATNF}
\end{table*}

\begin{table*}
\begin{tabular}{l|llllll}
\hline
JName      & $H$                    & $H^{(-0.8)}$     & $H^{(-0.4)}$     & $H^{(0)}$        & $H^{(0.4)}$     & $H^{(0.8)}$     \\
\hline
J0034$-$0721 & $0.5500\pm0.034$ & $0.3872\pm0.073$ & $0.5123\pm0.077$ & $0.5817\pm0.079$ & $0.6627\pm0.081$ & $0.8067\pm0.078$\\
J0055$+$5117 & $0.3749\pm0.021$ & $0.3833\pm0.073$ & $0.5143\pm0.078$ & $0.5790\pm0.081$ & $0.6589\pm0.083$ & $0.8120\pm0.080$\\
J0134$-$2937 & $0.4612\pm0.041$ & $0.3772\pm0.079$ & $0.5094\pm0.083$ & $0.5861\pm0.088$ & $0.6731\pm0.093$ & $0.8302\pm0.088$\\
J0141$+$6009 & $0.5327\pm0.030$ & $0.3917\pm0.051$ & $0.5099\pm0.055$ & $0.5741\pm0.059$ & $0.6514\pm0.057$ & $0.7969\pm0.056$ \\
J0147$+$5922 & $0.8283\pm0.027$ & $0.3847\pm0.060$ & $0.5103\pm0.068$ & $0.5923\pm0.069$ & $0.6703\pm0.073$ & $0.8164\pm0.069$ \\
J0151$-$0635 & $0.4837\pm0.093$ & $0.3870\pm0.110$ & $0.5185\pm0.120$ & $0.6097\pm0.130$ & $0.6938\pm0.130$ & $0.8522\pm0.130$\\
J0152$-$1637 & $0.5945\pm0.063$ & $0.3835\pm0.068$ & $0.5119\pm0.074$ & $0.5944\pm0.077$ & $0.6807\pm0.079$ & $0.8375\pm0.075$\\
J0215$+$6218 & $0.6041\pm0.043$ & $0.3847\pm0.069$ & $0.5085\pm0.073$ & $0.5783\pm0.075$ & $0.6554\pm0.080$ & $0.7971\pm0.077$ \\
J0304$+$1932 & $0.5465\pm0.063$ & $0.3802\pm0.067$ & $0.5044\pm0.073$ & $0.5750\pm0.075$ & $0.6472\pm0.076$ & $0.8025\pm0.078$ \\
J0357$+$5236 & $0.8939\pm0.025$ & $0.3847\pm0.057$ & $0.5121\pm0.063$ & $0.5840\pm0.064$ & $0.6678\pm0.065$ & $0.8204\pm0.064$ \\
J0502$+$4654 & $0.6647\pm0.036$ & $0.3879\pm0.060$ & $0.5090\pm0.064$ & $0.5815\pm0.065$ & $0.6630\pm0.069$ & $0.8067\pm0.065$\\
J0525$+$1115 & $0.5351\pm0.059$ & $0.3827\pm0.063$ & $0.5058\pm0.067$ & $0.5812\pm0.070$ & $0.6620\pm0.073$ & $0.8101\pm0.066$\\
J0601$-$0527 & $0.5020\pm0.037$ & $0.3983\pm0.080$ & $0.5063\pm0.081$ & $0.5609\pm0.089$ & $0.6291\pm0.093$ & $0.7625\pm0.093$\\
J0612$+$3721 & $0.7576\pm0.031$ & $0.3880\pm0.056$ & $0.5051\pm0.061$ & $0.5730\pm0.063$ & $0.6457\pm0.066$ & $0.7913\pm0.063$\\
J0630$-$2834 & $0.7764\pm0.026$ & $0.3912\pm0.045$ & $0.5098\pm0.049$ & $0.5800\pm0.050$ & $0.6519\pm0.052$ & $0.7991\pm0.050$\\
J0758$-$1528 & $0.7326\pm0.029$ & $0.3913\pm0.053$ & $0.5076\pm0.059$ & $0.5777\pm0.057$ & $0.6516\pm0.061$ & $0.7899\pm0.061$\\
J0814$+$7429 & $0.5972\pm0.031$ & $0.3938\pm0.054$ & $0.5076\pm0.056$ & $0.5708\pm0.059$ & $0.6352\pm0.059$ & $0.7753\pm0.058$\\
J0820$-$1350 & $0.5560\pm0.052$ & $0.3924\pm0.054$ & $0.5042\pm0.058$ & $0.5663\pm0.058$ & $0.6365\pm0.062$ & $0.7810\pm0.059$\\
J0823$+$0159 & $1.0916\pm0.030$ & $0.4322\pm0.130$ & $0.5432\pm0.140$ & $0.6033\pm0.150$ & $0.6642\pm0.160$ & $0.7727\pm0.170$\\
J0837$+$0610 & $0.6022\pm0.054$ & $0.3900\pm0.058$ & $0.5007\pm0.060$ & $0.5737\pm0.065$ & $0.6407\pm0.065$ & $0.7830\pm0.064$\\
J0837$-$4135 & $0.8023\pm0.061$ & $0.4057\pm0.100$ & $0.5270\pm0.110$ & $0.6020\pm0.110$ & $0.6701\pm0.120$ & $0.8085\pm0.120$\\
J0846$-$3533 & $0.6004\pm0.020$ & $0.3868\pm0.049$ & $0.5107\pm0.053$ & $0.5821\pm0.055$ & $0.6602\pm0.054$ & $0.8074\pm0.051$\\
J0908$-$1739 & $0.5353\pm0.043$ & $0.3875\pm0.064$ & $0.5080\pm0.074$ & $0.5760\pm0.074$ & $0.6552\pm0.073$ & $0.7975\pm0.072$\\
J0943$+$1631 & $0.5401\pm0.058$ & $0.3814\pm0.073$ & $0.5148\pm0.077$ & $0.5922\pm0.084$ & $0.6830\pm0.085$ & $0.8413\pm0.081$\\
J1022$+$1001 & $0.3625\pm0.043$ & $0.4268\pm0.100$ & $0.5277\pm0.110$ & $0.5802\pm0.120$ & $0.6448\pm0.120$ & $0.7727\pm0.130$\\
J1041$-$1942 & $0.5449\pm0.092$ & $0.4003\pm0.085$ & $0.5129\pm0.097$ & $0.5699\pm0.099$ & $0.6261\pm0.110$ & $0.7460\pm0.110$\\
J1115$+$5030 & $0.5736\pm0.071$ & $0.3905\pm0.077$ & $0.5102\pm0.088$ & $0.5703\pm0.090$ & $0.6463\pm0.090$ & $0.7851\pm0.092$\\
J1239$+$2453 & $0.5504\pm0.066$ & $0.4084\pm0.077$ & $0.5095\pm0.083$ & $0.5605\pm0.083$ & $0.6112\pm0.084$ & $0.7249\pm0.089$\\
J1257$-$1027 & $0.5695\pm0.057$ & $0.3839\pm0.068$ & $0.5077\pm0.074$ & $0.5796\pm0.075$ & $0.6499\pm0.078$ & $0.7999\pm0.073$\\
J1543$+$0929 & $0.5890\pm0.031$ & $0.3843\pm0.059$ & $0.5064\pm0.064$ & $0.5843\pm0.063$ & $0.6650\pm0.067$ & $0.8165\pm0.062$\\
J1703$-$3241 & $0.5996\pm0.049$ & $0.3912\pm0.057$ & $0.5099\pm0.063$ & $0.5715\pm0.062$ & $0.6485\pm0.061$ & $0.7880\pm0.063$\\
J1733$-$2228 & $0.5086\pm0.049$ & $0.3933\pm0.069$ & $0.5086\pm0.074$ & $0.5751\pm0.079$ & $0.6484\pm0.079$ & $0.7886\pm0.078$\\
J1741$-$0840 & $0.5158\pm0.070$ & $0.3860\pm0.082$ & $0.5153\pm0.088$ & $0.5840\pm0.094$ & $0.6725\pm0.098$ & $0.8264\pm0.089$\\
J1750$-$3157 & $0.5775\pm0.055$ & $0.3743\pm0.094$ & $0.5032\pm0.100$ & $0.5832\pm0.110$ & $0.6694\pm0.110$ & $0.8291\pm0.110$\\
J1756$-$2435 & $0.5530\pm0.086$ & $0.3795\pm0.100$ & $0.5120\pm0.110$ & $0.5847\pm0.120$ & $0.6703\pm0.120$ & $0.8234\pm0.120$\\
J1820$-$1818 & $0.5292\pm0.040$ & $0.3828\pm0.089$ & $0.5099\pm0.100$ & $0.5790\pm0.110$ & $0.6476\pm0.110$ & $0.8074\pm0.110$\\
J1822$-$2256 & $0.6462\pm0.028$ & $0.3826\pm0.075$ & $0.5088\pm0.084$ & $0.5925\pm0.090$ & $0.6767\pm0.084$ & $0.8320\pm0.079$\\
J1823$+$0550 & $0.5796\pm0.070$ & $0.3834\pm0.075$ & $0.5075\pm0.078$ & $0.5879\pm0.080$ & $0.6672\pm0.087$ & $0.8193\pm0.075$\\
J1834$-$0426 & $0.5913\pm0.019$ & $0.3845\pm0.097$ & $0.5116\pm0.110$ & $0.5839\pm0.110$ & $0.6610\pm0.120$ & $0.8057\pm0.120$\\
J1837$-$0653 & $0.5578\pm0.033$ & $0.3736\pm0.085$ & $0.5095\pm0.094$ & $0.5842\pm0.100$ & $0.6730\pm0.099$ & $0.8310\pm0.096$\\
J1840$+$5640 & $0.5495\pm0.040$ & $0.3848\pm0.057$ & $0.5054\pm0.061$ & $0.5776\pm0.065$ & $0.6550\pm0.063$ & $0.7976\pm0.061$\\
J1842$-$0359 & $0.3376\pm0.100$ & $0.4340\pm0.160$ & $0.5390\pm0.180$ & $0.6040\pm0.180$ & $0.6759\pm0.190$ & $0.8252\pm0.190$\\
J1852$-$2610 & $0.4759\pm0.037$ & $0.3807\pm0.087$ & $0.5111\pm0.096$ & $0.5829\pm0.099$ & $0.6737\pm0.100$ & $0.8229\pm0.100$\\
J1900$-$2600 & $0.4881\pm0.057$ & $0.3927\pm0.067$ & $0.5042\pm0.068$ & $0.5722\pm0.072$ & $0.6328\pm0.074$ & $0.7762\pm0.074$\\
J1901$-$0906 & $0.4590\pm0.069$ & $0.3787\pm0.091$ & $0.5126\pm0.097$ & $0.5911\pm0.110$ & $0.6725\pm0.100$ & $0.8306\pm0.110$\\
J1921$+$2153 & $0.5835\pm0.056$ & $0.3906\pm0.055$ & $0.5063\pm0.059$ & $0.5793\pm0.059$ & $0.6565\pm0.063$ & $0.8064\pm0.058$\\
J1946$+$1805 & $0.5013\pm0.038$ & $0.3912\pm0.065$ & $0.5038\pm0.071$ & $0.5700\pm0.073$ & $0.6339\pm0.074$ & $0.7699\pm0.075$\\
J1954$+$2923 & $0.4457\pm0.097$ & $0.3859\pm0.110$ & $0.5106\pm0.120$ & $0.5822\pm0.130$ & $0.6664\pm0.130$ & $0.8076\pm0.130$\\
J2018$+$2839 & $0.6467\pm0.051$ & $0.3881\pm0.053$ & $0.5038\pm0.056$ & $0.5752\pm0.058$ & $0.6462\pm0.062$ & $0.7925\pm0.059$\\
J2046$-$0421 & $0.4396\pm0.078$ & $0.3773\pm0.092$ & $0.5101\pm0.098$ & $0.5934\pm0.110$ & $0.6838\pm0.110$ & $0.8398\pm0.100$\\
J2046$+$1540 & $0.5234\pm0.080$ & $0.3764\pm0.094$ & $0.5105\pm0.100$ & $0.5968\pm0.110$ & $0.6855\pm0.110$ & $0.8472\pm0.110$\\
J2108$+$4441 & $0.5712\pm0.026$ & $0.3865\pm0.049$ & $0.5087\pm0.051$ & $0.5797\pm0.057$ & $0.6532\pm0.053$ & $0.7972\pm0.052$\\
J2113$+$4644 & $0.6291\pm0.027$ & $0.3867\pm0.051$ & $0.5061\pm0.053$ & $0.5764\pm0.056$ & $0.6476\pm0.058$ & $0.7967\pm0.054$\\
J2308$+$5547 & $0.4547\pm0.043$ & $0.3892\pm0.071$ & $0.5030\pm0.076$ & $0.5710\pm0.078$ & $0.6508\pm0.080$ & $0.7945\pm0.080$\\
\hline
\end{tabular}
\caption{Hurst parameters for timing noise series of $54$ pulsars
together with 5 artificial (anti-)persistent series for comparison.
See \S \ref{sec_mc} for explanation of symbols.}
\label{tab_H}
\end{table*}

\end{document}